\documentstyle[preprint,aps]{revtex}

\begin{document}

\bibliographystyle{prsty} 

\draft 
\input epsf.sty 

\author{Kevin K. Lehmann}
\address{Department of Chemistry, Princeton University, 
Princeton NJ 08544 USA}

\date{June 10, 2000}

\title{Rotation in liquid $^4$He: Lessons from a toy model}

\maketitle

\begin{abstract}
 This paper presents an analysis of a model
problem, consisting of two interacting rigid rings,
for the rotation of molecules in liquid $^4$He.  
Due to Bose symmetry, the excitation of the
rotor corresponding to a ring of N helium atoms
is restricted to states with integer multiples of N
quanta of angular momentum.  This minimal model
shares many of the same features of the rotational
spectra that have been observed for molecules in 
nanodroplets of $\approx 10^3 - 10^4$ helium atoms.
In particular, this model predicts, for the first time, the very large
enhancement of the centrifugal distortion constants
that have been observed experimentally.  It also illustrates the different
effects of increasing rotational velocity by increases
in angular momentum quantum number or by increasing the
rotational constant of the molecular rotor.  
It is found that fixed node, diffusion Monte Carlo and
a hydrodynamic model provide upper and lower bounds on the
size of the effective rotational constant of the molecular
rotor when coupled to the helium.

\end{abstract}

\newpage

The spectroscopy of atoms and molecules dissolved in helium
nanodroplets is a topic of intense current
interest~\cite{Toennies98,Grebenev00b,Lehmann98}. One particular, almost unique
feature of this spectroscopic host is that even heavy and very anisotropic
molecules and complexes give spectra with rotationally resolved
structure~\cite{Hartmann96}. This spectral structure typically
corresponds to thermal equilibrium, with $T\approx 0.38$\,K,
and has the same
symmetry as that of the same species in the gas
phase~\cite{Hartmann95,Callegari00b}. The rotational constants, however, are
generally reduced by a factor of up to four or five, while the
centrifugal distortion constants are four orders of magnitude 
larger than for the gas phase~\cite{Grebenev00a,Callegari00b}.
These large changes clearly reflect dynamical coupling between
the molecular rotation and helium motion.  At present, there are
at least four different models proposed for the increased effective moments
of inertia, at least two of which have reported quantitative 
agreement with experiment~\cite{Grebenev98,Kwon99b,Callegari99b,Gordon00}.  The
large observed distortion constants have not yet been quantitatively explained,
and the most careful attempt to date to calculate them (for OCS in helium) gave
an estimate
$\approx 30$ times smaller than the experimental value~\cite{Grebenev00a}

The highly quantum many body dynamics of this condensed phase system has
made it difficult to achieve a qualitative understanding of the
observed effects.  In cases like these, simple models
can provide insight, especially
if the lessons learned can be tested against more computationally
demanding simulations that seek, however, to provide a first principles
treatment of the properties of the system of interest.
In this paper, one very simple model system will be explored
that seeks to model the coupling of a molecular rotor to
a first solvation shell of helium.  The existing models for
the reduced rotational constants agree that most of the observed effect
comes from motion of helium in the first solvation shell.
Some of the qualitative features of this model were discussed
previously~\cite{Grebenev00a}, but quanitative details were not persued in
that work.

The `toy' model considered consists of a planar rotor coupled to a symmetric
planar ring of $N$ helium atoms.  This model problem can be solved
exactly, and can reproduce the size of the observed reductions in
the rotational constant AND the size of the centrifugal distortion 
constants.  This is the first time, to the authors knowledge,
that the large effective distortion constants of molecules 
in liquid helium has been reproduced.
Further, this model clearly resolves a confusion about the {\em sign} of the
centrifugal distortion constant.  
Based upon the expected decreased following of the helium with
increasing rotational angular velocity~\cite{Lee99,Conjusteau00}, one
can argue that the rotational spacing should increase faster than for a
rigid rotor,{\em i.e.}, that the effective centrifugal distortion constant
should be {\em negative}, in conflict with experimental observations.  The
present model demonstrates, however, that opposite behavior
is expected when the rotational velocity of the rotor is increased by increasing
the rotational quantum number (where an increased angular anisotropy and
following of the helium is predicted) or when the rotational constant of the
isolated rotor is increased (where decreased angular anisotropy and following of
the helium is predicted).  The present model, therefore, rationalizes both the
observed depenence of the increased moments of inertia on the rotational
constant of the isolated molecule and the observed centrifugal distortion
constants.

\section{The toy model}

We will consider a highly abstracted model for rotation of a
molecule in liquid helium.  The molecule will be treated as
a rigid, planar rotor with moment of inertia $I_1$. 
The orientation of the molecule is given by $\theta_1$
The liquid helium is treated as a ring of $N$ helium atoms
that forms another rigid, planar rotor with moment of inertia $I_2$ and with
orientation given by $\theta_2$. Because of the Bose symmetry of the helium, he
helium rotor can only be excited to states with $N \hbar$ units of angular
momentum.  The lowest order symmetry allowed coupling between the molecule
and the helium ring is given by a
potential $V\cos \left[ N(\theta_1 - \theta_2)\right]$.
Any coupling spectral components that are not multiples of $N$ will
lead to mixing of states that are not allowed by Bose symmetry, which is
forbidden in quantum mechanics.
The Hamiltonian is given by:
\begin{equation}
H = - \frac{\hbar^2}{2 I_1} \frac{\partial^2}{\partial \theta_1^2}
 - \frac{\hbar^2}{2 I_2} \frac{\partial^2}{\partial \theta_2^2}
+ V \cos N(\theta_1 - \theta_2) \label{eq:Huc}
\end{equation}
We define $B_{1,2} = \frac{\hbar^2}{2 I_{1,2}}$, the rotational constants
for the uncoupled rotors.
We can separate the above $H$ by introducing the two new coordinates:
\begin{equation}
\bar{\theta} = \frac{I_1 \theta_1 + I_2 \theta_2}{I_1 + I_2} \;\;\;\;\;\;\;
\theta = \theta_1 - \theta_2 \label{eq:Hc}
\end{equation}
in which we have:
\begin{equation}
H = H_{\rm r} + H_{\rm v} = - \frac{\hbar^2}{2 (I_1 + I_2)}
\frac{\partial^2}{\partial
\bar{\theta_1}^2} + \left[
- \frac{\hbar^2}{2} \left(\frac{1}{I_1}+\frac{1}{I_2}\right)
\frac{\partial^2}{\partial \theta^2} + V \cos (N\theta) \right]
\end{equation}
$\bar{\theta}$ is the variable conjugate to the total angular
momentum; $\theta$ is a vibrational coordinate.  We define $B_{\rm rigid} =
\frac{\hbar^2}{2 (I_1 + I_2)}$ and
$B_{\rm rel} = \frac{\hbar^2}{2} \left(\frac{1}{I_1}+\frac{1}{I_2}\right)$.
The eigenstates of $H$ separate into a product:
\begin{equation}
\psi(\bar{\theta},\theta) = e^{i J \bar{\theta}} \psi_{\rm
v}(\theta)
\end{equation}
$J$ is the quantum number for total angular momentum.
It would appear from the separable $H$ that the energy could
be written as an uncoupled sum of a rigid rotor energy, $B_{\rm rigid} J^2$
( not $J(J+1)$ because we have a planar rotor), and
a `vibrational' energy that is independent of $J$.  However, the energies are
{\em not} simply additive, due to the fact that the boundary condition for
$\theta$ is $J$-dependent.  
When $\theta_2$ alone is changed by any multiple of $2\pi/N$, $\psi$
must be unchanged.  However, a change of $-2\pi/N$ in $\theta_2$
results in a change of $-(2\pi/N)(I_2/(I_1 + I_2)$ in $\bar{\theta}$
and $+2\pi/N$ in $\theta$.  Thus, the Bose symmetry of the helium
ring is satisfied by taking as the boundary condition for $\psi_{\rm v}$:
\begin{equation}
\psi_{\rm v} \left(\theta + \frac{2\pi}{N} \right) =
\exp \left( \frac{2\pi i}{N} \frac{I_2}{I_1 + I_2} \, J \right) \,
\psi_{\rm v} (\theta) \label{eq:bc1}
\end{equation}
As a result, the `vibrational energies' and eigenfunctions are a function
of the total angular momentum quantum number, $J$.
Note that the boundary condition is periodic in $J$, with
period given by $N (I_1 + I_2)/ I_2 $.

The $J$ dependence of the boundary condition of the vibrational
function is rather unfamiliar in molecular physics.  This
dependence can be removed by a Unitary transformation of
the wavefunction:
\begin{equation}
\psi'_{\rm v} (\theta) = \exp \left( - i \frac{I_2 J}{I_1 + I_2} \theta \right)
\psi_{\rm v} (\theta)
\end{equation}
The boundary condition on the transformed function is
$\psi'_{rm v}  \left(\theta + \frac{2\pi}{N} \right) = 
\psi'_{rm v}  \left(\theta \right)$.  The
transformed hamiltonian, $H'$, is given by:
\begin{equation}
H' = \frac{1}{2 I_1} \left( J - L \right)^2 + \frac{1}{2 I_2} L^2
+ V \cos \left( N \theta \right)
\end{equation}
where $J = i \hbar \frac{\partial}{\partial \bar{\theta}}$ is
the operator for total angular momentum, and 
$L = i \hbar \frac{\partial}{\partial \theta}$ is the
angular momentum of the helium ring relative to
a frame moving with the rotor.  This form of the
Hamiltonian closely resembles those widely used
in treatment of weakly  bound complexes~\cite{Hutson91}.
Expansion of the $\left( J - L \right)^2$ gives
a Coriolis term that couples the overall rotation
to the vibrational motion, which makes the
nonseperability of these motions evident.
This form, however, hides the periodicity in $J$ of this
the coupling.

We will now consider the two limiting cases.  The first to consider
is that where $|V| \gg N^2 B_{\rm rel}$.  In this case, the potential
can be considered harmonic around $\theta = 2\pi k/N, k=0 \ldots
N-1$ when $V$ is negative (and shifted by $\pi/N$ for $V$ positive) with
harmonic frequency
$\nu = \frac{1}{2 \pi}
\sqrt{2 N^2 |V|\left(\frac{1}{I_1}+\frac{1}{I_2}\right)}$.  The wavefunction
will decay to nearly zero at the maxima of the potential, and changes
in the phase of the periodic boundary condition at this point (which
happens with changes in $J$) will not significantly affect the energy.  In
this limit, the total energy is $E(J,{\rm v}) = B_{\rm rigid} J^2 + h\nu
({\rm v} + 1/2)$, and we have a rigid rotor spectrum with effective moment
of inertia $I_1 + I_2$. 
While there are $N$ equivalent minima, Bose symmetry assures that only
one linear combination of the states localized in each well (the
totally symmetric combination for $J=0$) is allowed, and thus there are
no small tunneling splittings, even in the high barrier limit.
The total angular momentum is partitioned between the
two rotors in proportion to their moments of inertia, {\it i.e.} $<J_1> =
\hbar J
\cdot I_1 / (I_1 + I_2)$ and $<J_2> = \hbar J \cdot I_2 /(I_1 + I_2)$.

We will now consider the opposite, or uncoupled rotor, limit.  The
eigenenergies in this case are trivially $E(m_1,m_2) = B_1 m_1^2 + B_2
m_2^2$ with eigenfunctions $\psi = \exp \left( i m_1 \theta_1 + i m_2
\theta_2 \right)$. $m_1$ can be any integer, while $m_2 = N k$,
where $k$ is any integer. Introducing the total angular momentum quantum
number $J = m_1 + m_2$, we have $E(J,m_2) = B_1 (J - m_2)^2 +
B_2 \,m_2^2$.    
The lowest state for each $J$ has quantum numbers $m_1 = J - Nk$ and
$m_2 = Nk$, where $k$ is the nearest integer to 
$B_1 J/N(B_1 +B_2)$.   
Treating the quantum numbers as continuous, we have $E = B_{\rm rigid} J^2$,
{\it e.g.}, the same as for rigid rotation of the rotors. 
However when we restrict
$J$ to integer values, for $J \le N(B_1 + B_2)/(2 B_1)$, the energy
spacing will be exactly that of a rigid rotor with rotational constant
$B_1$.  In general, as a function of $J$, the uncoupled ground state
solutions follow the rigid rotor spectrum in $B_1$, but with a series of
equally spaced curve crossings when the lowest energy $m_2$ value increases
by $N$ as $J$ is increased by one quantum.
These crossings allow the total energy to oscillate around that predicted
for a rigid rotor with moment of inertia $I_1 + I_2$.

\section{Numerical Results}

Having handled the limiting cases, we can now turn our attention
to the far more interesting question, which is how does the energy
and eigenstate properties change as $V$ is continuously varied between
these limits.  We note that changing the sign of $V$ is equivalent to
translation of the solution by $\Delta\theta = \pi$, and thus we
will only consider positive values of $V$ explicitly.  We also note
that the eigenstates are not changed, and the eigenenergies
scale linearly if $B_1, B_2$, and $V$ are multiplied by a constant factor.  
As a result, we will take $B_2 =1$ to normalize the energy scale.
The solutions for finite values of $V$ were calculated using the
uncoupled basis and the form of $H$ given in Eq.~\ref{eq:Huc} with fixed
values of
$m_1 + m_2 = J$ and $m_2 = Nk$.
For each value of
$J$, the matrix representation for $H$ is a tridiagonal matrix, with
diagonal elements given by the energies for the uncoupled limit, and with
off-diagonal elements given by $V/2$.  Numerical calculations were
done using a finite basis with $k = -15,-14,\ldots 15$.

\begin{figure}
\centerline{\epsfbox{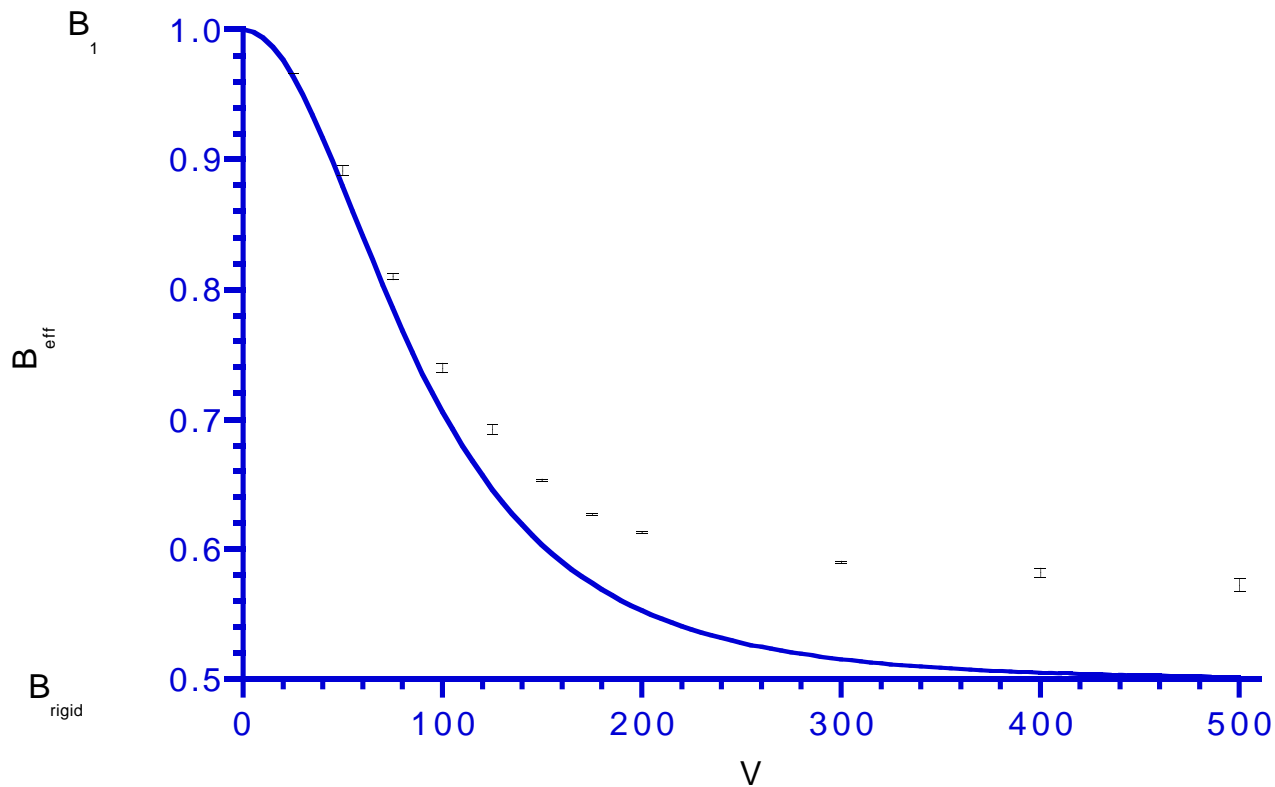}}
\vspace{1in}
\caption{Effective Rotational Constant for two coupled rotors as a function
of the interaction potential strength. 
$B_1 = B_2 = 1$, and the second rotor can only be excited to states with
multiples of 8 quanta.  The individual points are the $B$ effective values
calclated by a fixed node, Diffusion Monte Carlo calculation.  The error
bars on these points are the estimated $2\sigma$ sampling error.}
\label{figBeff}
\end{figure}

Using $B_1 = B_2$ and $N=8$, we have calculated the lowest eigenvalues of $H$
for $J=0,1,2$ and used these,
by fitting to the expression $E({\rm v},J) = E_0({\rm v})
+ B_{\rm eff} J^2 - D_{\rm eff} J^4$,
to determine $B_{\rm eff}$ and $D_{\rm eff}$.  
Figure~\ref{figBeff} shows the value of $B_{\rm eff}$ as a 
function of $V$ (both in units of $B_2$).
It can be seen that $B_{\rm eff}$ varies smoothly from $B_1$
to $B_{\rm rigid}$ with increasing $V$, and is reaches a value
half way between these limits for $V \approx N^2 B_2$.

In order to rationalize this observation, we will now consider 
a Quantum Hydrodynamic treatment for the rotation~\cite{Leggett73}.
Let the ground state density be $\rho(\theta) = 
|\psi_{\rm v} (\theta)|^2$.  Now let the molecule classically rotate
with  angular velocity $\omega$.  To first order
in $\omega$, $\rho$ will not change (i.e. we will have
adiabatic following of the helium density
for classical infinitesimal rotation of the molecule).  However,
the vibrational wavefunction, $\psi_{\rm v}$ will no longer be real, but
instead will have an angle-dependent phase factor whose
gradient will give a hydrodynamic velocity.
Solving the equation of continuity:
\begin{equation}
\frac{\rm d}{{\rm d}\theta} \left( \rho v \right) = 
-\frac{{\rm d}\rho}{{\rm d}t} = 
\omega r \frac{{\rm d}\rho}{{\rm d}\theta} 
\end{equation}
where $r$ is the radius of the helium ring, gives
solutions of the form:
\begin{equation}
v(\theta) = \omega r \cdot \left( 1 - \frac{C}{\rho(\theta)} \right)
\end{equation}
where $C$ is an integration constant.  We determine $C$ by
minimizing the kinetic energy averaged over $\theta$.  This
gives:
\begin{equation}
C = \frac{2\pi}{\int_{0}^{2\pi} \rho^{-1}}
\end{equation}
and a kinetic energy:
\begin{equation}
\Delta E_{k} = \frac{1}{2} I_2 \omega^2 \cdot 
\left( 1 - \frac{4\pi^2}
{\int_{0}^{2\pi} \rho^{-1} {\rm d} \theta}  \right) 
\end{equation}
In the case of a uniform density, $\rho = (2\pi)^{-1}$ and 
$\Delta E_{k} = 0$.  
As the density gets more anisotropic, the integral becomes larger
and $\Delta E_{k}$ becomes larger, approaching the value
for rigid rotation of the helium ring when $\rho$ has a node
in its angular range.  We define the hydrodynamic
contribution to the increase in the moment of inertia of
the heavy rotor due to partial rotation of the light rotor
by $\Delta E_{k} =  \frac{1}{2} \Delta I_{\rm h} \omega^2$.
It is interesting to note that for the above lowest energy value of $C$,
we have $\int_0^{2\pi} v(\theta) {\rm d}\theta = 0.$ ({\it i.e.}
that the solution is `irrotational') and that
the net angular momentum induced in the helium is
$\Delta I_{\rm h} \omega$.  The lowest energy solution of the three dimensional
Quantum hydrodynamic model satifies these conditions as
well~\cite{Callegari99b,Callegari00a}.

\begin{figure}
\centerline{\epsfbox{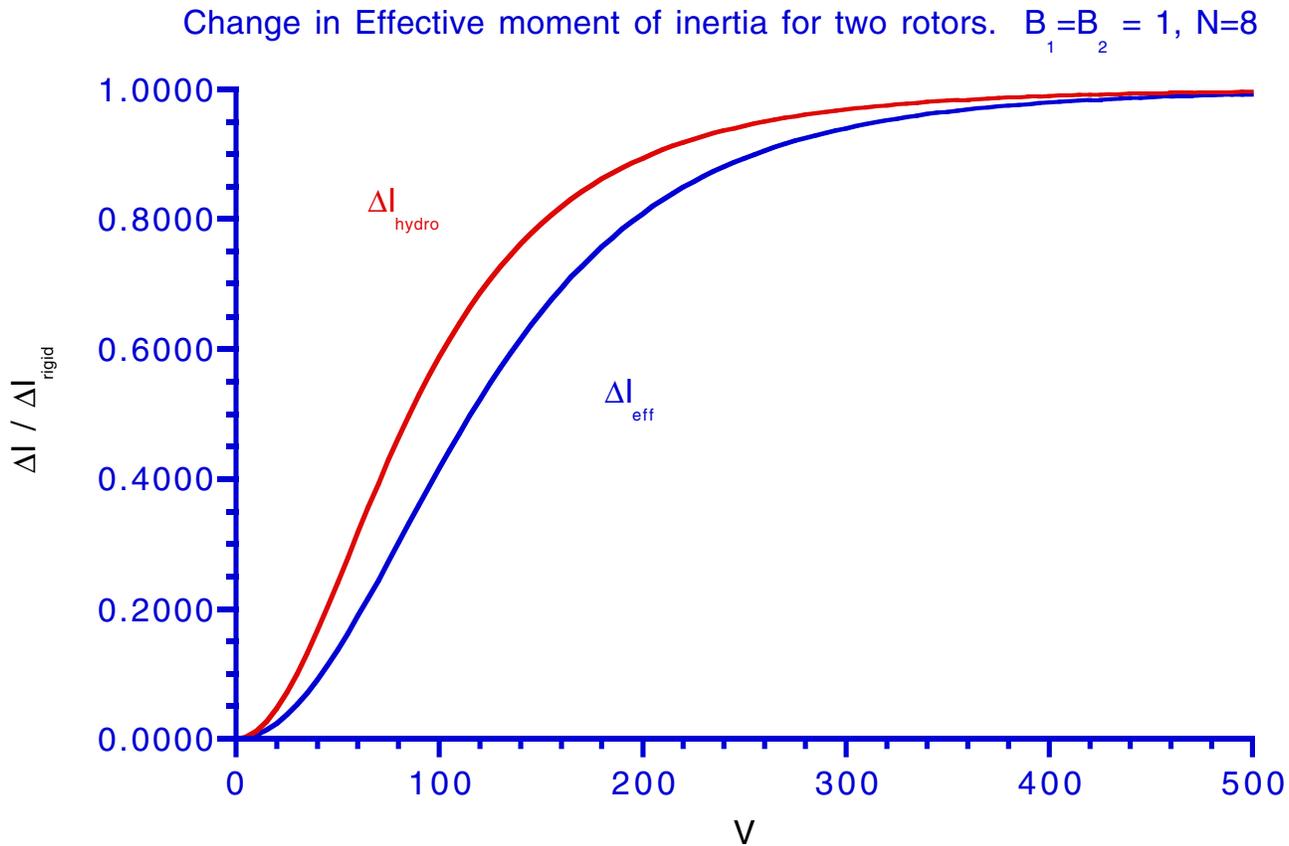}}
\vspace{1in}
\caption{Increase in effective moment of inertia of molecule, $\Delta_{\rm
eff}$ due to coupling to rotor made of 8 helium atoms. $\Delta I_{\rm hydro}$
is the same quantity estimated by the hydrodynamic model.  Both are
calculated as a function of the potential coupling strength, and the results
normalized to the rigid rotor moment of inertia of the 8 helium rotor}
\label{figDeltaIeff}
\end{figure}

The hydrodynamic model can be tested against the exact quantum solutions.
Define $\Delta I_{\rm eff}$ as the effective moment of
inertia for rotation (as calculated from $B_{\rm eff}$) minus the
moment of inertia for the molecular rotor. 
$\Delta I_{\rm eff}$ will grow from $0$ for uncoupled
rotors to $I_2$ as the coupling approachs the rigid coupling limit
of high $V$. 
In the hydrodynamic model, 
$\Delta I_{\rm eff} = \Delta I_{\rm h}$.
 Figure~\ref{figDeltaIeff} shows a plot that compares
$\Delta I_{\rm eff}$ and $\Delta I_{\rm h}$ 
as a function of $V$.  Each has been normalized
by $I_2$.   They are found to be in qualitative agreement
for the full range of $V$, though the exact quantum solution
is systematically below the hydrodynamic prediction.
We note, however, that for the assumed parameters, the
speed of the molecular rotor is equal to that of the 
helium rotor, while the hydrodynamic treatment assumed
a classical, infinitesimal rotation of the molecular rotor.  

\begin{figure}
\centerline{\epsfbox{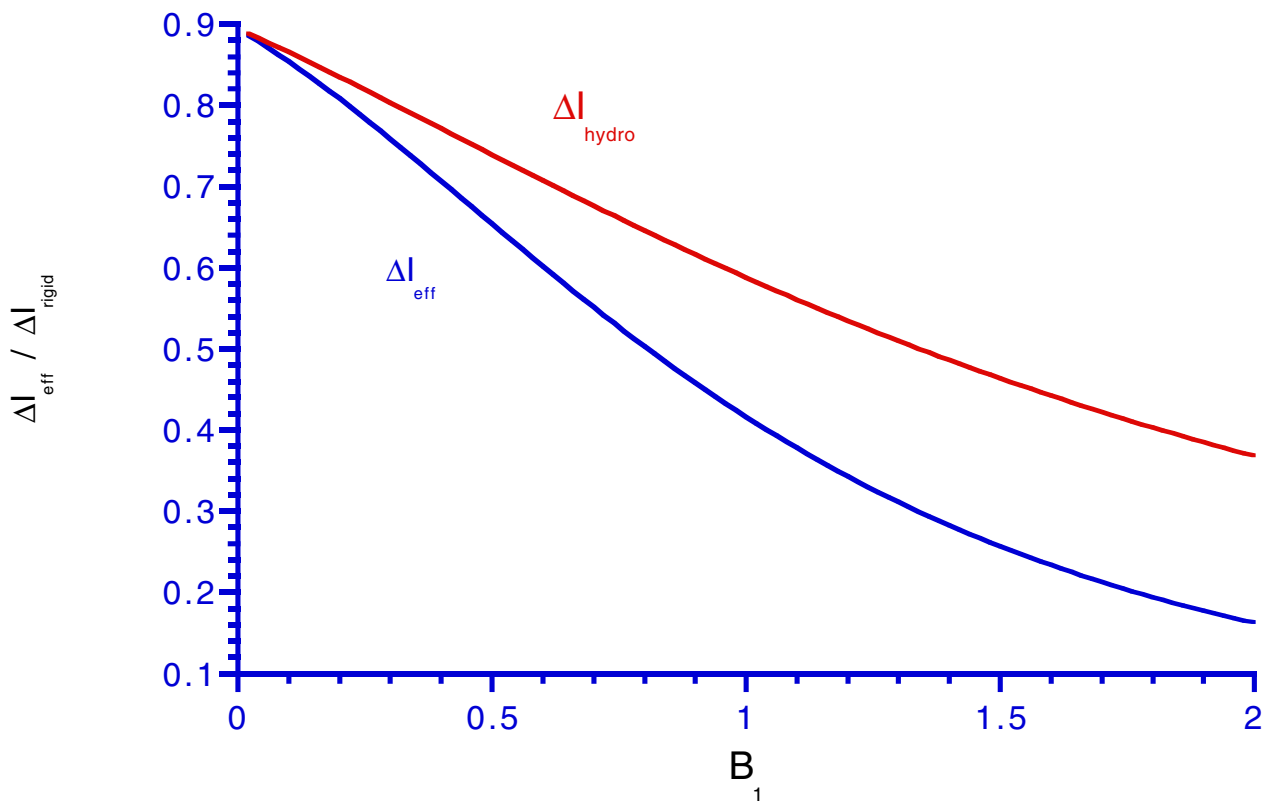}}
\vspace{1in}
\caption{Same as Figure~\ref{figDeltaIeff}, except as a function of the
rotational constant of the molecule, normalized to the rotational constant
of the 8 helium rotor.}
\label{figB1}
\end{figure}

The size of $\Delta I_{\rm eff}$ is determined by the 
degree of anisotropy of the ground state density in
the vibrational displacement coordinate $\theta$.
If $I_1$ is decreased at fixed $I_2$ and $V$, the 
effective mass for $\theta$, which is
$(I_1^{-1} + I_2^{-1})^{-1}$ will also decrease,
which will decrease the anisotropy produced by $V$.
Fig~\ref{figB1} shows how the normalized
$\Delta I_{\rm eff}$ and $\Delta I_{\rm h}$ vary as
the molecular rotational constant,
$B_1$, changes from $0$ to $2 B_2$.
This calculation was done for $V = 100$, close to
the value corresponding to maximum difference of 
$\Delta I_{\rm eff}$ and $\Delta I_{\rm h}$ for $B_1 = B_2$.
This plot demonstrates that the hydrodynamic prediction
becomes exact in the limit that $B_1 \rightarrow 0$,
{\it i.e.}, in the case that the assumption of
infinitesimal rotational velocity of the molecule holds. 
However, it substantially overestimates the increase
effective moment of inertia when $B_1 \ge B_2$.
This decrease in the increase moment of inertia with
increasing rotational constant of the heavy rotor is
the effect previously interpreted as the breakdown of adiabatic following
in the literature on the rotational spectrum of
molecules in liquid Helium~\cite{Lee99,Callegari99b,Conjusteau00}.

\begin{figure}
\centerline{\epsfbox{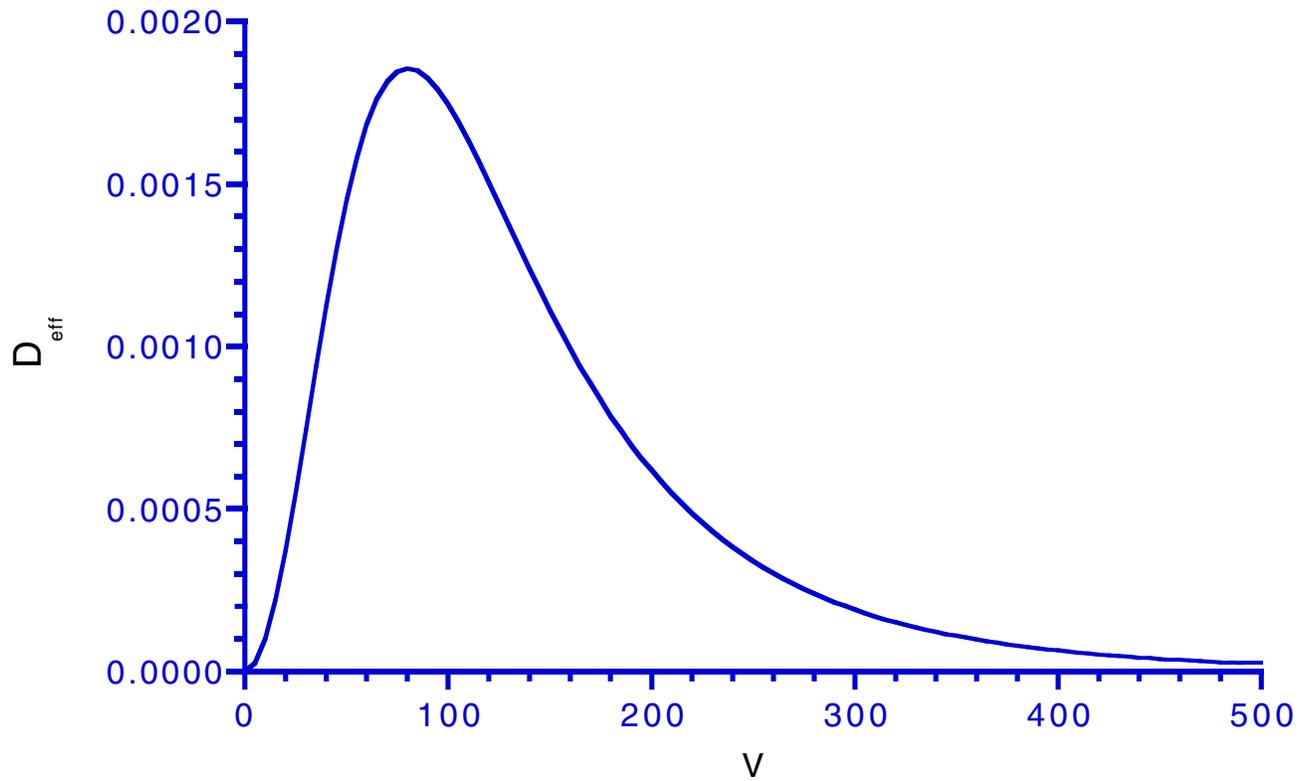}}
\vspace{1in}
\caption{The effective centrifugal distortion constant, $D_{\rm eff}$, for
molecule coupled to ring of 8 helium atoms as a function of the strength of
the coupling, $V$.  Both $D_{\rm eff}$ and $V$ are normalized to the
rotational constants of the molecule and 8 helium rotor, which are taken as
equal.}
\label{figdeff}
\end{figure}

Figure~\ref{figdeff} shows a plot of $D_{\rm eff}$ as a function
of $V$ for $B_1 = B_2 =1$.  $D_{\rm eff} = 0$ is zero in both limits, and has
a maximum value near the value of $V$ at which $B_{\rm eff}$ is changing most
rapidly.  
It is interesting to explicitly point out that this 
$D_{\rm eff}$ value arises entirely from changes in the angular
anisotropy of the helium density with $J$, 
as the model does not allow for an increase in the
radial distance of the helium, which has previously
been considered~\cite{Grebenev00a}.  
Further, the peak value of $D_{\rm eff} \approx 1.8\,\cdot 10^{-3} B_1$ is in
remarkably good agreement with the ratio of $D_{\rm eff}$ to the gas phase
molecular rotational constant observed for a number of molecules in liquid
helium.  For example, for OCS this ratio is found to be
$2 \cdot 10^{-3}$~\cite{Grebenev00a},  while for HCCCN, the same ratio was found
to be $1 \cdot 10^{-3}$~\cite{Callegari00c}.

\begin{figure}
\centerline{\epsfbox{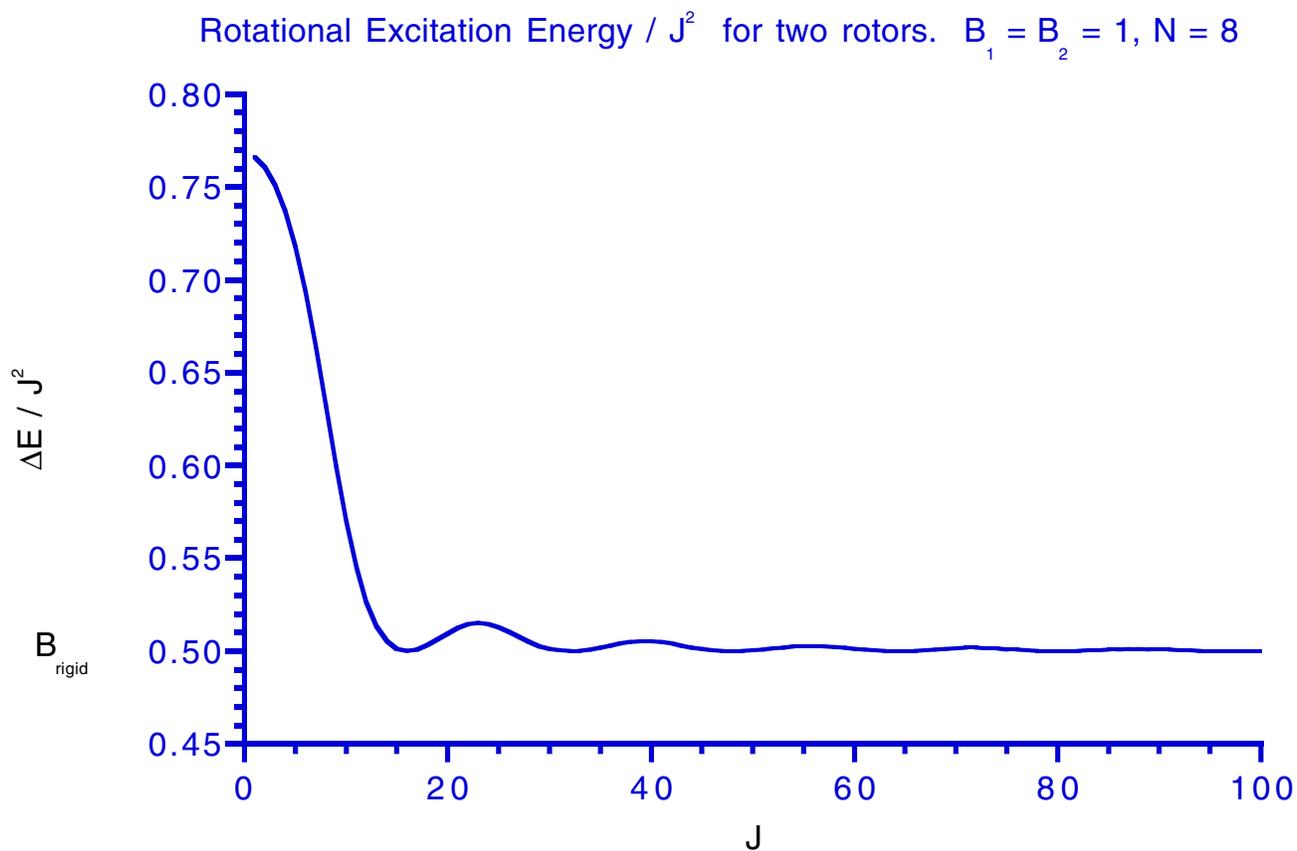}}
\vspace{1in}
\caption{The rotational excitation energy, $\Delta E$, divided by $J^2$ as
a function of the total rotational angular momentum quantum number, $J$. 
Calculated with $B_1 = B_2 = 1$ and $V = 100$.  With rigid following of the
helium, the plotted quantity should equal $B_{\rm rigid}$, which is
indicated in the figure.}
\label{figDeltaE}
\end{figure}

\begin{figure}
\centerline{\epsfbox{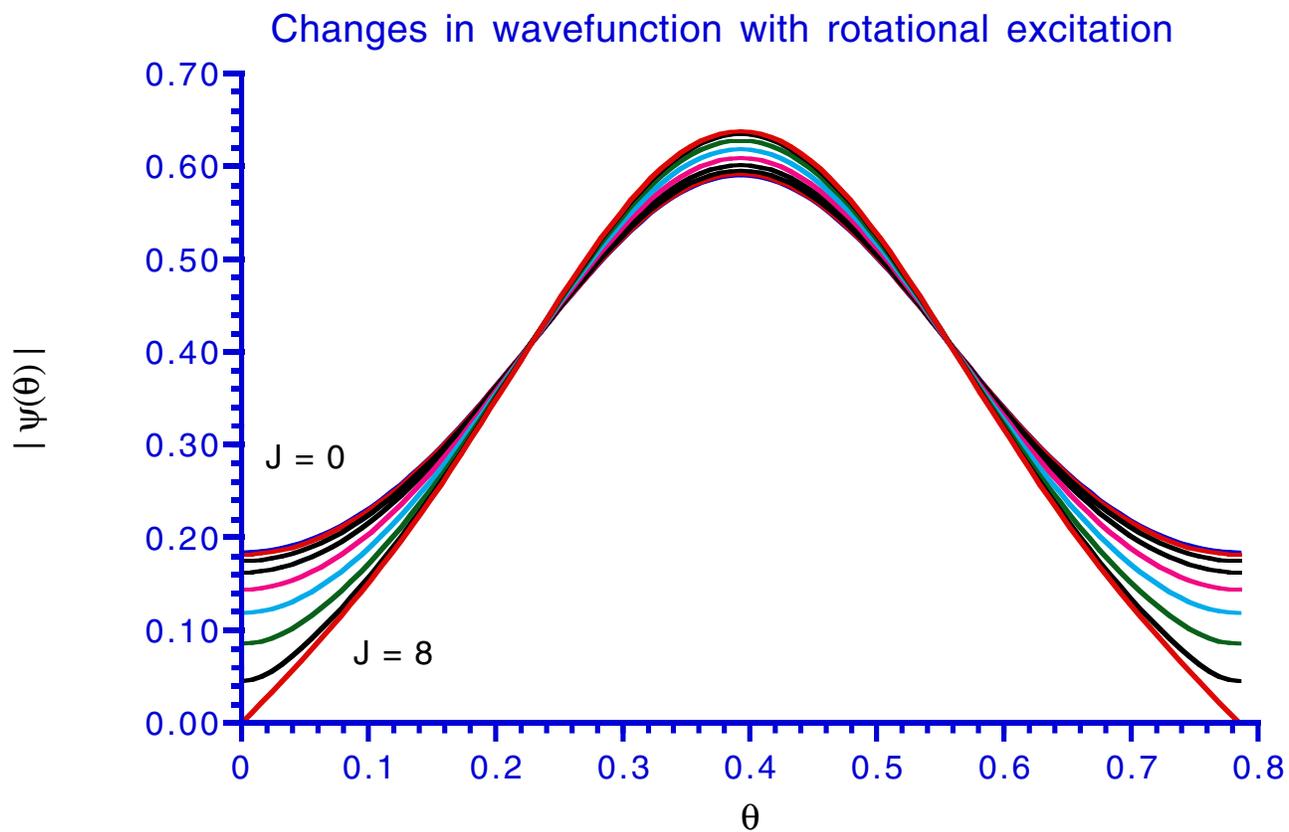}}
\vspace{1in}
\caption{The absolute value of the vibrational wavefunction as a
function of the relative orientation between molecule and 8 helium
rotor.  This quantity is plotted for angular momentum $J=0,1\ldots8$.}
\label{figpsi}
\end{figure}

We can gain further insight by examining the rotational
energy systematically as a function of $J$.  
Figure~\ref{figDeltaE} shows the rotational excitation
energy ($E(0,J)-E(0,0)$) divided by $J^2$ as a function of $J$.
The calculations were done with $V = 100$.
The rotational excitation energy approaches that of the $B_{\rm rigid} J^2$
for high $J$.  
Further, it reaches this value for $J$ equal to multiples of
$N(I_1 + I_2)/I_2$, which matches the   
periodicity of the boundary conditions for $\psi_{\rm v}$.
$J$ values that lead to the same boundary conditions for $\psi_{\rm v}$
will differ in energy only by the eigenvalues of $H_{\rm r}$, and thus
it follows from Eq.~\ref{eq:Hc},
that of a rigid rotor with rotational constant $B_{\rm rigid}$.
For the first
half of each period in $J$, $\psi_{\rm v}$ is found to increase in its
anisotropy, and therefore the energy increases, as $J$ is increased (See
Fig.~\ref{figpsi}).  This can be understood when one considers the fact that for
$J = N(I_1+I_2)/(2I_2)$, the boundary condition is that
$\psi_{\rm v}(2\pi/N) = - \psi_{\rm v}(0)$, i.e. the wavefunction
will be real but have $N$ nodes in the interval $[0,2\pi]$.  

Classically, the molecular rotor is characterized by its rotational
angular velocity, $\omega = 2 B_1 J$.  However, we see that the
quantum treatment of the two coupled rotors gives opposite results
when $\omega$ is increased by increasing either $B_1$ or $J$. 
For increases in $B_1$, the `degree of following' of the 
light rotor decreases for fixed potential coupling, as seems
intuitively reasonable.  However, for increases in $J$,
the anisotropy of the potential and thus the `degree of
following' initially increases, and thus so does the effective
moment of inertia of the coupled system.  This behavior
continues until one passes through a resonance condition
where the helium can be excited by transfer of $N$ quantum of
angular momentum from molecular rotor to the helium.  
This resonance condition is missing from the classical treatment
of the coupling between the rotors, where the angular velocity 
of the molecular rotor is treated as
a fixed quantity, $\omega$, which is one of the parameters of the
problem. 

\section{Nodal Properties of Solutions}

It is possible to calculate the rotational excitation energies
of clusters of helium around a molecule by use of the Fixed Frame Diffusion
Monte Carlo (FFDMC) method~\cite{Lee99}.  
As in most DMC methods, this method should yield (except for
statistical fluctuations) a upper bound on the true energy, finding
the optimal wavefunction consistent with the nodal properties that
are imposed on the wavefunction by construction.  
In the case of FFDMC, the nodal planes are determined by the free rotor
rotational wavefunction for the molecule alone, i.e. that the sign
of the wavefunction (which is taken to be real) for any point in
configuration space is the same as that of the rotor wavefunction at
the same Euler angles.

We can examine the exact solutions of our toy problem to gain
insight into the accuracy of the nodal planes assumed in FFDMC.
The wavefunctions we have considered up to now are complex, but
because of time reversal symmetry, the solutions with $J$ and $-J$
rotational quantum numbers must be degenerate.  Symmetric combination
of these solutions just gives the Real part of $J$ solution, and
the antisymmetric combination the Imaginary part.  The real part
is given by:
\begin{equation}
\psi^{\rm R}_J (\theta_1,\theta_2) =
\cos(J \theta_1) Re(\psi'(\theta_2 - \theta_1))
- \sin(J \theta_2) Im(\psi'(\theta_2 - \theta_1))
\end{equation}
where
\begin{equation}
\psi'(\Delta \theta) = \sum_k c^J_k \exp(i k N \Delta\theta)
\end{equation}
and $c^J_k$ are the eigenvector coefficients obtained from diagonalization
of the real Hamiltonian matrix in the uncoupled basis.
Examination of the numerical solutions reveals that for $J < N$, 
$Re(\psi')$ has no nodes, while $Im(\psi')$ has nodes at $\Delta\theta$
equal to integer multiples of $\pi/N$.  
Thus, if $Im(\psi') = 0$, then the solution would satisfy the 
FFDMC nodal properties exactly.  However, for finite $Im(\psi')$,
the nodal surfaces, rather than being on the planes $\theta_1 =$\,constant,
are modulated N times per cycle along the $\theta_1 =$\,constant line.
For $V=80$ and $B_1 = B_2 = 1$, the maximum value of $Im(\psi')$ is about
$4\%$ of $Re(\psi')$, and growing approximately linearly for low $J$.

In order to test the quantiative implications of this error in
the nodal properties, importance sampled DMC calculations have been
done for the present two rotor problem.  The explicit DMC algorithm
given by Reynolds {\it et al.}~\cite{Reynolds82} was used with minor
change~\cite{DMC_note}.   The guiding function, $\psi_{\rm T}$, which determines
the nodes, was selected as $\cos( J \theta_1) \psi_{\rm v}^{0} (\theta_1 -
\theta_2)$, where $\psi_{\rm v}^{0}$ is the real, positive definite eigenstate
for the $J=0$ problem.  The rotational constant, $B_{\rm DMC}$ is defined
as the DMC estimated energy for $J=1$ less the exact ground state
eigenenergy for $J=0$, and will be (except for sampling and finite
time step bias) an upper bound on the true $B$ value calculated earlier.
The points plotted in figure~\ref{figBeff} are the calculated values of
$B_{\rm DMC}$ with the estimated $2\sigma$ error estimates.  It is
seen that the fixed node DMC estimates of $B_{\rm eff}$ are excellent
for low values of $V$, but underestimate the contribution of the Helium
ring to the effective moment of inertia as it is coupled more strongly to the
rotor.

\section{Relationship with a more Realistic Model}

In a series of insightful lectures, Anthony Leggett analyzed the
properties of the ground state of $N$ Helium atoms confined to an annulus
of radius $R$ and spacing $d \ll R$~\cite{Leggett73}.  
The walls of the annulus are allowed to classically
rotate with angular velocity $\omega$.
While not stated explicitly, the walls of the annulus couple to
the helium via a time dependent potential, which is
static in the rotating frame. 
As such, our rotating diatomic molecule can be considered as
a special case of the problem treated by Leggett.
If one transforms to the rotating frame, the
quantum hamiltonian is the same as for the static ($\omega = 0$)
problem.  However, the boundary condition for the wavefunction
in this frame is given by \cite{Leggett73}[Eq. (2.10)]:
\begin{equation}
\Psi'_0(\theta_1,\theta_2\ldots\theta_j+2\pi\ldots\theta_N) =
\exp \left( 2 \pi i m R^2 \omega /\hbar \right)
\Psi'_0(\theta_1,\theta_2\ldots\theta_j\ldots\theta_N) \label{eq:bc2}
\end{equation}
In making a comparison to the results of the toy model, we note that
for this system
$I_2 = N m R^2$ (the classical moment
of inertia for the helium) and $J = \omega (I_1 + I_2)/\hbar$.
Substitution shows that the phase factor in
Eq.~\ref{eq:bc2} is identical to that derived above for Eq.~\ref{eq:bc1}.
Note, however, that Eq.~\ref{eq:bc2} refers to moving one helium atom by
$2\pi$, while Eq.~\ref{eq:bc1} refers to motion of all $N$ helium atoms
by $2\pi/N$.  Motion of all $N$ helium atoms by $2\pi$ will result in
a phase factor of $2 \pi i N m R^2 \omega /\hbar = 2\pi i I_2 J / (I_1 +
I_2)$ in both treatments.

Leggett considered the change in helium energy produced by rotation
of the walls.
Let $E_0$ be the ground state energy for the static problem, and
$E'_0(\omega)$ the ground state energy in the rotating frame.
The ground state energy in the laboratory frame is given by
\cite{Leggett73}[Eq. (2.12)]:
\begin{equation}
E_{\rm lab} = E_0 + \frac{1}{2}I_2 \omega^2 - 
\left[E'_0 (\omega) - E_0 \right]
\end{equation}
For the ground state of Bosons, we further have that 
$E'_0 (\omega) \ge E_0$, with equality only when
$\omega$ equals integer multiples of $\omega_0 = \hbar/m R^2$
since the nodeless state has the lowest possible energy.
At $\omega = k\omega_0$, the helium rigidly rotates with
the walls.  
This agrees exactly with the numerical
results of the toy model, as shown in Figure~\ref{figDeltaE}.
In making comparisons with this model, one should remember that
$E_{\rm lab}$ does not include the kinetic energy of the walls (rotor).
Thus the more general treatment of Leggett supports one of the central
insights of the toy model, that the large effective distortion constants for
molecular rotors in helium is a consequence of an {\em increased}
helium following of the rotor with increasing angular velocity,
which in turn is a direct consequence of the $\omega$ dependence
of the single-valuedness boundary condition in the rotating frame.

The moment of inertia for the ground state of the helium
can be defined by:
\begin{equation}
I = \left( \frac{d^2 E_{\rm lab}}{d\omega^2} \right)_{\omega \rightarrow 0}
= I_2 - \left( \frac{d^2 E'_0(\omega)}{d\omega^2} \right)_{\omega
\rightarrow 0}
\end{equation}
Leggett defined the ``normal fraction'' of the helium by
the ratio $I/I_2$, which is equal to unity if 
$E'_0(\omega)$ is independent of $\omega$ as $\omega \rightarrow 0$.
This will occur if the wavefunction has `nontrivial' nodal planes,
since the phase of the wavefunction can be changed discontinuously at
a node without cost of energy.  
Nodal plans associated with overlap of particles, however, are
`trivial' in that the phase relationship on each side of the node is
determined by the exchange symmetry of the wavefunction, and thus
cannot be used to match the boundary conditions without extra cost of
energy.  In our toy problem, when $V$ is very large, the vibrational
wavefunction becomes localized, introducing near nodes at the maxima
of the potential, and as a result the ground state is described by
a near unity normal fraction; we have what Leggett refers to as a
`normal solid'.  
Conversely, as the uncoupled limit is approached, the
helium ring does not contribute to the kinetic energy of the lowest
rotational states and we have $I \rightarrow 0$, and we have
zero normal fraction (i.e. the helium has unity superfluid fraction).  
Following Leggett's definition, one finds that the normal fraction
is given by $\Delta I_{\rm eff}/I_2$.  Thus,
Figure~\ref{figDeltaIeff} can thus be interpreted as the
normal fluid fraction for the ground state as a function
of the strength of the potential coupling.  
Leggett's analysis is based upon a classical treatment of the
motion of the walls, which implies $I_1 \gg I_2$, in
which limit the hydrodynamic model exactly predicts
the normal fluid fraction.

\section{Acknowledgement}

This work was supported by the National Science Foundation
and the Air Force Office of Scientific Research.  

\bibliography{Helium}

\begin{thebibliography}{10}
\bibitem{DMC_note}
It was found that when Walkers moved very close to a node, they became trapped,
  due to the large size of the attempted steps introduced by the drift term
  combined with the `detailed balance' correction selection criteria. When they
  also had large negative values of the local energy, these trapped walkers
  then grew in weight, leading to unphysical large negative values for the DMC
  estimate for the energy. This problem, which is part of the finite time step
  bias, was eliminated by killing walkers that failed the detail balance
  selection instead of keeping them at their previous location.
\end{thebibliography}

\end{document}